\documentclass[10pt]{iopart}
\usepackage{graphicx}
\usepackage{color}

\begin{document}

%\title[Universality of quantum criticality in Yb-based heavy fermions]{Universality of quantum criticality in Yb-based heavy fermions on quasicrystal and pressurized approximant crystal}
\title[Effects of Crystalline Electronic Field and Onsite Interorbital Interaction in Yb-based Quasicrystal and AC]{Effects of Crystalline Electronic Field and Onsite Interorbital Interaction in Yb-based Quasicrystal and Approximant Crystal}

\author{Shinji Watanabe$^1$ and Kazumasa Miyake$^2$}

\address{$^1$Department of Basic Sciences, Kyushu Institute of Technology, Kitakyushu, Fukuoka 804-8550, Japan \\
$^2$Center for Advanced High Magnetic Field Science, Osaka University, Toyonaka 560-0043, Japan}
%\ead{submissions@iop.org}
\vspace{10pt}
\begin{indented}
\item[]December 2017
\end{indented}

\begin{abstract}
To get an insight into a new type of quantum critical phenomena recently discovered in the quasicrystal Yb$_{15}$Al$_{34}$Au$_{51}$ and approximant crystal (AC) Yb$_{14}$Al$_{35}$Au$_{51}$ under pressure, 
we discuss the property of the crystalline electronic field (CEF) at Yb in the AC 
and show that uneven CEF levels at each Yb site can appear because of the Al/Au mixed sites.   
Then we construct the minimal model for the electronic state on the AC
by introducing the onsite Coulomb repulsion between the 4f and 5d orbitals at Yb. 
Numerical calculation for the ground state shows that  
the lattice constant dependence of the Yb valence well explains the recent measurement done by systematic substitution of elements of Al and Au in the quasicrystal and AC,   
where the quasicrystal Yb$_{15}$Al$_{34}$Au$_{51}$ is just located at the point from where the Yb-valence starts to change drastically. Our calculation convincingly demonstrates that this is indeed the evidence that this material is just located at the quantum critical point of the Yb-valence transition.  
\end{abstract}

%\maketitle
\ioptwocol

\section{Introduction}

Quantum critical phenomena have attracted great interest in condensed matter physics. Quantum critical phenomena emerging near the magnetic quantum critical point (QCP) have been well understood from the spin fluctuation theory by Moriya~\cite{Moriya} and the renormalization group theory by Hertz~\cite{Hertz} and Millis~\cite{Millis}.
However, a new type of quantum criticality, which does not follow the conventional critical spin fluctuation theory, has been discovered in heavy-electron metals such as YbCu$_{5-x}$Al$_x$~\cite{Bauer}, YbRh$_2$Si$_2$~\cite{Trovarelli} and $\beta$-YbAlB$_4$~\cite{Nakatsuji}. 
As a possible origin, the theory of critical Yb-valence fluctuation 
%%%%%
%\textcolor{red}
%{
(CVF) 
%}
%%%%%
has been proposed by the present authors~\cite{WM2010}, which gives a unified explanation for the unconventional critical phenomena observed in a series of physical quantities. 

Interestingly, the common unconventional quantum criticality has been discovered in the heavy electron quasicrystal Yb$_{15}$Al$_{34}$Au$_{51}$~\cite{Deguchi,Watanuki}. 
This is the first quasicrystal with the intermediate valence of Yb among Yb-based quasicrystals, although the valence of Yb was Yb$^{2+}$ in the quasicrystal including Yb so far. On the other hand, the X-ray spectroscopy measurement has revealed Yb$^{+2.66}$ at $T=300$~K in Yb$_{15}$Al$_{34}$Au$_{51}$~\cite{Watanuki}.
This material exhibits the non-Fermi liquid behaviors in a series of physical quantities such as the magnetic susceptibility $\chi\sim T^{-0.5}$, the specific-heat coefficient $C/T\sim -\ln{T}$, and the resistivity $\rho\sim T$ at ambient pressure and zero magnetic field. 
Surprisingly, the criticality persists even under pressure, at least up to $P=1.6$~GPa~\cite{Deguchi}. Namely, the quantum criticality is quite robust against pressure. 
Furthermore, a new type of scaling called ``$T/B$ scaling" has been observed in the quasicrystal, where 
the magnetic susceptibility $\chi$ can be expressed as a single scaling function of the ratio of the temperature $T$ and a magnetic field $B$ over 6 decades~\cite{Matsukawa2016,Deguchi2017}. 
This $T/B$ scaling is essentially the same as that observed in $\beta$-YbAlB$_4$ with the same scaling function~\cite{Matsumoto}.

%%%%%%%%%%%%  Fig.1  %%%%%%%%%%%%%%%%%%%%%%%%%%%%%%
\begin{figure*}
\includegraphics[width=15cm]{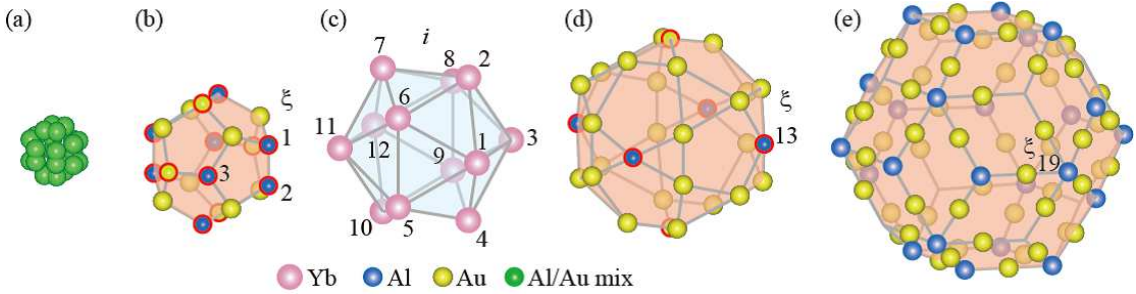}
\caption{(Color online) Tsai-type cluster, which consists of the concentric shell structures of (a)1st shell, (b)2nd shell, (c)3rd shell, (d)4th shell, and (e)5th shell. In (b) and (d), the sites framed in red indicate the Al/Au mixed sites. 
The number $i$ specifies the Yb site in (c) and $\xi$ specifies the Al site in (b), (d), and (e). 
} 
\label{fig:YbAuAl}
\end{figure*}
%%%%%%%%%%%%%%%%%%%%%%%%%%%%%%%%%%%%%%%%%%%%%%%%%%%

To clarify the mechanism of these emergent phenomena, theoretical studies have been performed by the present authors~\cite{WM2013,WM2015,WM2016}. 
Figure~\ref{fig:YbAuAl} illustrates the Tsai-type cluster, which is the core structure of the quasicrystal Yb$_{15}$Al$_{34}$Au$_{51}$. 
The Tsai-type cluster consists of the concentric shell structures shown in Figs.~\ref{fig:YbAuAl}(a)-\ref{fig:YbAuAl}(e)~\cite{Deguchi,Ishimasa}. 
There also exists the 1/1 approximant crystal (AC) Yb$_{14}$Al$_{35}$Au$_{51}$. 
The AC has the periodic arrangement of the Tsai-type cluster, which forms the body-center-cubic (bcc) lattice. 
In the 3rd shell [Fig.~\ref{fig:YbAuAl}(c)], the 12 atoms of Yb are located at the vertices of the icosahedron. 
In the 1st shell [Fig.~\ref{fig:YbAuAl}(a)], the 2nd shell [Fig.~\ref{fig:YbAuAl}(b)], and the 4th shell [Fig.~\ref{fig:YbAuAl}(d)], Al/Au mixed sites exist with existence ratio 7.8\%/8.9\%, 62\%/38\%, and 59\%/41\%, respectively. 
Such an example of atomic distributions following the existence ratios is illustrated in Fig.~\ref{fig:YbAuAl}. 

Theoretical analysis of the electronic state of the extended Anderson model (with the Coulomb repulsion between the 4f and conduction electrons $U_{\rm fc}$) on the Tsai-type cluster has shown that many spots of the QCP of the Yb-valence transition appears in the ground-state phase diagram in the $U_{\rm fc}$-$\varepsilon_{\rm f}$ plane with $\varepsilon_{\rm f}$ being the energy level of the 4f hole of Yb ion because of the difference of the f-c hybridization at each Yb site due to the Al/Au mixed sites [see Figs.~\ref{fig:YbAuAl}(b) and \ref{fig:YbAuAl}(d)]~\cite{WM2013}. Hence, the quantum critical regions are overlapped and unified, giving rise to the vast critical region in the phase diagram. 
Since the infinite limit of the unit-cell size of the AC corresponds to the quasicrystal, the quantum critical region in the quasicrystal is much wider than that of the AC~\cite{WM2015}. 
This gives a natural explanation for the robust criticality under pressure and also zero-tuning criticality in the quasicrystal~\cite{WM2013,WM2015}.

Furthermore, theoretical analysis of the periodic Anderson model of the AC, where the Tsai-type cluster is periodically arranged on the bcc lattice, has shown that the pressurized AC exhibits the quantum criticality of Yb-valence fluctuation in the magnetic susceptibility as $\chi\sim T^{-0.5}$ for zero magnetic-field limit and the $T/B$ scaling behavior as observed in the quasicrystal~\cite{WM2016}.    
The key origin has been clarified to be the locality of the 
%%%%%
%\textcolor{red}
%{
CVF
%}
%%%%%
~\cite{WM2014}. 
%%%%%
%\textcolor{red}
%{
Namely, almost dispersionless CVF mode appears in the momentum space, giving rise to the extremely small characteristic temperature of CVF $T_0$. In the temperature region for $T/T_0>1$, a new type of quantum criticality emerges in physical quantities~\cite{WM2010,WM2014}. 
%}
%%%%%
Since as the unit-cell size of the AC increases, the locality is expected to be further enhanced, 
%%%%%
%\textcolor{red}
%{
i.e., $T_0$ is expected to become further smaller, 
%}
%%%%%
 this mechanism also explains the unconventional criticality and the $T/B$ scaling observed in the quasicrystal as mentioned above~\cite{WM2016}. Actually, recent measurement in the AC has detected that $\chi\sim T^{-0.5}$ and the $T/B$ scaling, which are the same as those observed in the quasicrystal, appear under pressure $P=1.96$~GPa~\cite{Matsukawa2016}.

In this paper, we further develop the theory on the Yb-based quasicrystal and AC. We focus on the following two aspects, which have not been studied before: One is the effect of the crystalline electronic field (CEF) and the other is the effect of the onsite interorbital Coulomb repulsion. 
By analyzing the CEF in the AC, we find that the energy level of the 4f ground state can be different at each Yb site because of the presence of the Al/Au mixed sites. This further supports the previous result~\cite{WM2013,WM2015} that difference in the f-c hybridization at each Yb site gives rise to the wide quantum critical region. 
By constructing the extended periodic Anderson model with the onsite interorbital Coulomb repulsion, we will show that the valence QCP is realized with the intermediate Yb valence for realistic values of parameters. 
We also find that the lattice constant dependence of the Yb valence well explains the recent experiments done by systematic substitution of the elements Al and Au in the quasicrystal Yb-Al-Au and AC. 

The organization of this paper is as follows. In Sect.~2, we discuss the nature of the CEF in the AC and the quasicrystal. In Sect.~3, we construct the extended periodic Anderson model for the AC by introducing the onsite interorbital Coulomb repulsion and discuss the ground-state property on the basis of the numerical results. The paper is summarized in Sect.~4.

%----------------------------------------------------------------------------------------------
\section{Crystalline Electronic Field of the Yb-Al-Au Approximant Crystal}

\begin{figure}[h]
\includegraphics[width=5cm]{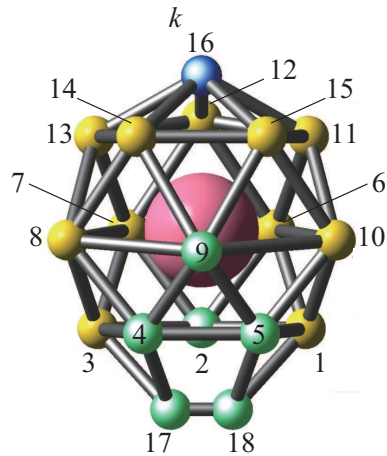}
\caption{(Color online) Local configuration around the Yb atom (pink) (see Fig.~\ref{fig:YbAuAl})~\cite{Matsukawa2014}.  The number $k$ labels surrounding Al (blue), Au (yellow), and Al/Au mixed site (green).} 
\label{fig:Yb_local}
\end{figure}

In this section, we discuss the property of the crystalline electronic field (CEF) of the 4f electron at the Yb site. 
The space group of the 1/1 AC is $Im\bar{3}$ No.204 and the site symmetry at the Yb site is $m$~\cite{Ishimasa}. This mirror symmetry holds under an assumption that the uniform distribution of the same atoms on the 1st shell and Al/Au mixed sites on the 2nd and 4th shells in Fig.~\ref{fig:YbAuAl}. 
The local configuration around the Yb atom was discussed by Matsukawa {\it et al}.,~\cite{Matsukawa2014}, as shown in Fig.~\ref{fig:Yb_local}. This configuration is realized at each Yb site in the AC and also 70~\% of Yb atoms in the quasicrystal~\cite{Watanuki,Matsukawa2014}. 

In Fig.~\ref{fig:Yb_local}, the number $k\ (=1, \cdots 18)$ specifies the surrounding atoms of Yb, occupying the Al, Au, and Al/Au mixed site. 
To see the relation to the Tsai-type cluster shown in Fig.~\ref{fig:YbAuAl}, 
for instance, if we focus on the $i=1$st Yb site in  Fig.~\ref{fig:YbAuAl}(c), $\xi=1$st, 2nd, and 3rd sites on the 2nd shell [Fig.~\ref{fig:YbAuAl}(b)] correspond to the $k=5$th, 4th, and 2nd sites in Fig.~\ref{fig:Yb_local}, respectively. The $\xi=13$th site on the 4th shell [Fig.~\ref{fig:YbAuAl}(d)] corresponds to the $k=9$th site in Fig.~\ref{fig:Yb_local}. The $\xi=19$th Al site on the 5th shell [Fig.~\ref{fig:YbAuAl}(e)] corresponds to the $k=16$th Al site in Fig.~\ref{fig:Yb_local} and the Au atoms surrounding the $\xi=19$th site forming the pentagon in Fig.~\ref{fig:YbAuAl}(e) corresponds to the Au atoms labeled by $k=11\sim 15$ in Fig.~\ref{fig:Yb_local}. The Al/Au mixed sites on the 1st shell [Fig.~\ref{fig:YbAuAl}(a)] correspond to the $k=17$th and 18th sites in Fig.~\ref{fig:Yb_local}.

%%%%%%%%%%%%  Fig.2  %%%%%%%%%%%%%%%%%%%%%%%%%%%%%%
%%%%%%%%%%%%%%%%%%%%%%%%%%%%%%%%%%%%%%%%%%%%%%%%%%%

As a first step of analysis of the CEF, we discuss the nature on the basis of the point charge model. 
The CEF Hamiltonian is given by
$H_{\rm CEF}=|e|V_{\rm cry}(\vec{r})$. Since the 4f$^{14}$ configuration is the closed shell and here we consider the 4f$^{13}$ state, we take the hole picture for $H_{\rm CEF}$, where the charge of the 4f hole at Yb is set as $|e|$. The potential $V_{\rm cry}(\vec{r})$ is given by 
\begin{eqnarray}
V_{\rm cry}(\vec{r})&=&\sum_{k=1}^{18}\frac{q_k}{\left|\vec{R}_k-\vec{r}\right|},
\nonumber
\\
&=&\sum_{k=1}^{18}\sum_{\ell=0}^{\infty}\sum_{m=-\ell}^{\ell}\frac{q_k}{R_k}\left(\frac{r}{R_k}\right)^{\ell}
\frac{4\pi(-1)^m}{2\ell+1}Y_{m}^{\ell}(\theta_k, \varphi_k)
\nonumber
\\
& &\times
Y_{-m}^{\ell}(\theta, \varphi),
\label{eq:Vcry}
\end{eqnarray}
where $Y^{m}_{\ell}(\theta,\varphi)$ is the spherical harmonics with the azimuthal quantum number $\ell$ and the magnetic quantum number $m$. Here, $\vec{R}_k$ specifies the atom position of the Al, Au, and Al/Au mixed site surrounding Yb, whose position is set to be the origin, as shown in Fig.~\ref{fig:Yb_local}. In the usual case where the lattice structure possesses the inversion symmetry, $V_{\rm cry}(-\vec{r})=V_{\rm cry}(\vec{r})$, the parity of the CEF eigenstate can be classified as ether even or odd. This makes the odd-$\ell$ term in Eq.~(\ref{eq:Vcry}) vanish and $V_{\rm cry}(\vec{r})$  can be expressed as the Stevens operators for the even-$\ell$ term with $\ell\le 6$~\cite{Stevens,Hutchings}. 
However, in the present case, $V_{\rm cry}(\vec{r})$ does not possess the inversion symmetry even if all the Al/Au sites are occupied by Au, as shown in Fig.~\ref{fig:Yb_local}, e.g., the $k=6\sim 10$ sites forms the pentagon, the vertex of which violates the inversion symmetry each other around the Yb site.  
Then, the odd-$\ell$ terms appear in Eq.~(\ref{eq:Vcry}), which cannot be expressed by the Stevens operators. 
Hence, we discuss the qualitative features of the CEF below.

From the analysis of the even-$\ell$ terms in Eq.~(\ref{eq:Vcry}) expressed by the Stevens operators with the use of the coordinates of the atom positions reported in Ref.~\cite{Ishimasa}, it turns out that $H_{\rm CEF}$ for the $|J_{z}\rangle$ states $(J_z=-7/2, -5/2, ..., 5/2, 7/2)$ for in the $J=7/2$ manifold has non-zero values of most of off-diagonal elements in addition to the diagonal elements. 
By diagonalizing the $8\times 8$ $H_{\rm CEF}$ matrix, the eigenvalues are split into four levels 
and the ground state is the Kramers doublet 
\begin{eqnarray}
|\Psi_{+}\rangle&=&a_{1}\left|\frac{7}{2}\right\rangle+a_{2}\left|\frac{5}{2}\right\rangle+a_{3}\left|\frac{3}{2}\right\rangle+a_{4}\left|\frac{1}{2}\right\rangle
\nonumber
\\
&+&a_{5}\left|-\frac{1}{2}\right\rangle+a_{6}\left|-\frac{3}{2}\right\rangle+a_{7}\left|-\frac{5}{2}\right\rangle+a_{8}\left|-\frac{7}{2}\right\rangle,
\label{eq:WFup}
\\
|\Psi_{-}\rangle&=&a_{8}^{*}\left|\frac{7}{2}\right\rangle-a_{7}^{*}\left|\frac{5}{2}\right\rangle+a_{6}^{*}\left|\frac{3}{2}\right\rangle-a_{5}^{*}\left|\frac{1}{2}\right\rangle
\nonumber
\\
&+&a_{4}^{*}\left|-\frac{1}{2}\right\rangle-a_{3}^{*}\left|-\frac{3}{2}\right\rangle+a_{2}^{*}\left|-\frac{5}{2}\right\rangle-a_{1}^{*}\left|-\frac{7}{2}\right\rangle, 
\label{eq:WFdw}
\end{eqnarray}
where $a_i$ is the complex numbers satisfying $\sum_{i=1}^{8}|a_i|^2=1$. 
Since non-zero matrix elements of $H_{\rm CEF}$ are expected not to be altered even after taking account of the odd-$\ell$ terms in Eq.~(\ref{eq:Vcry}), the CEF ground state of $H_{\rm CEF}$ is  considered to have the form as Eqs.~(\ref{eq:WFup}) and (\ref{eq:WFdw}) in general. 

%%%%%%%%%%%%  Fig.3  %%%%%%%%%%%%%%%%%%%%%%%%%%%%%%
\begin{figure}[h]
\includegraphics[width=7cm]{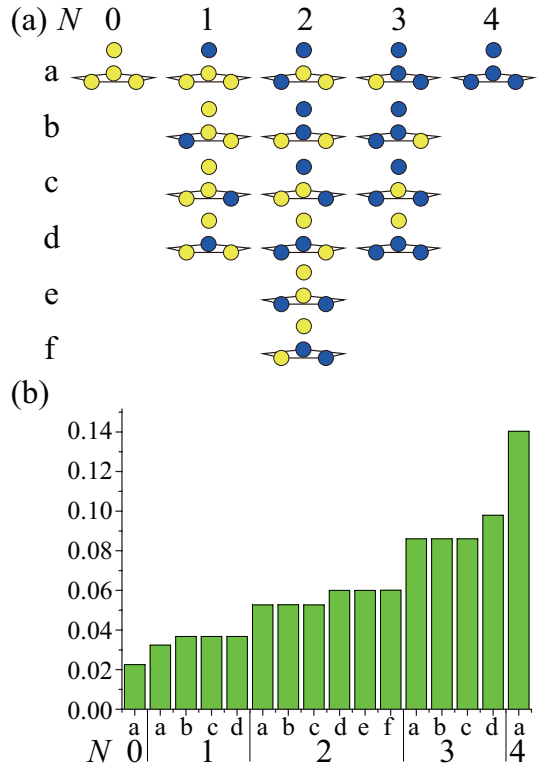}
\caption{(Color online) (a) Atomic configurations for the Al/Au mixed sites classified by the number of Al atoms. Al (blue) and Au (yellow) atoms. (b) Rate of each configuration of (a).} 
\label{fig:rate}
\end{figure}
%%%%%%%%%%%%%%%%%%%%%%%%%%%%%%%%%%%%%%%%%%%%%%%%%%%

The CEF energy of the ground state has various values depending on the charge distributions around the Yb site because of the existence of the Al/Au mixed sites. 
To see this, in Fig.~\ref{fig:rate}(a), 
we illustrate possible configurations of atoms at the Al/Au mixed sites specified by $k=2$, 4, 5, and 9 in Fig.~\ref{fig:Yb_local}. Here we neglect the contribution from the 1st shell due to the relatively small existence ratio (Al/Au: 7.8~$\%$/8.9~$\%$). The atomic configurations are classified by the number $N$ of Al atoms
as $N=0$~(1), 1~(4), 2~(6), 3~(4), and 4~(1), where the number in (\ ) indicates that of configurations. 
Since the Al ion and Au ion have the different charges in general, each configuration in Fig.~\ref{fig:rate}(a) gives the different CEF energies not only for the ground state but also for the excited states. 

The difference in the charges of the Al and Au ions can be quantified as follows: Since the unit cell of the AC contains 24 Yb, 62 Al, and 90 Au atoms in total~\cite{Ishimasa}, the condition for the neutrality of the total charge 
%%%%%
%\textcolor{red}
%{
%
\begin{eqnarray}
24Z_{\rm Yb}|e|+62Z_{\rm Al}|e|+90Z|e|=0, 
\label{eq:charge}
\end{eqnarray}
where the charges of Yb, Al and Au ions are expressed as $Z_{\rm Yb}|e|$, $Z_{\rm Al}|e|$ and $Z|e|$, respectively. 
Then, the charge of the Al ion is expressed as $-(45Z+12Z_{\rm Yb})|e|/31$. By inputting $q_k=Z|e|$ at the Au site and $q_k=-(45Z+12Z_{\rm Yb})|e|/31$ at the Al site in Eq.~(\ref{eq:Vcry}), the CEF energy can be obtained as a function of $Z_{\rm Yb}$ and $Z$.
%}
%%%%%

To estimate the rate of each configuration shown in Fig.~\ref{fig:rate}(a), we performed numerical calculation where the Al or Au atom is distributed randomly at the Al/Au mixed sites so as to follow the existence ratio mentioned above. The result of the calculation for the $10^7$ step of the random-number generation is shown in Fig.~\ref{fig:rate}(b). The vertical axis shows the rate of each configuration and the sum total gives 1.0. 
The CEF energy levels to be observed experimentally, e.g., neutron measurement, is expected to correspond to the averaged over the CEF energy levels for each configuration with the weight factor by multiplying each rate shown in Fig.~\ref{fig:rate}(b).

An important result here is that the CEF ground-state energy can be different at each Yb site depending on the surrounding atomic configuration shown in Fig.~\ref{fig:rate}(a).  
In the previous theoretical studies on the AC and quasicrystal~\cite{WM2013,WM2015}, it was shown that the effective f-c hybridization at each Yb site can be different because of the Al/Au mixed sites, which gives rise to the many spots of the valence QCP in the ground-state phase diagram. 
The result clarified here further assists the previous result since uneven f-level energies at each Yb site promote the tendency of causing the valence QCP at various locations in the phase diagram.  

In the quasicrystal, 70$~\%$ of Yb is located in the local configuration shown in Fig.~\ref{fig:Yb_local} and the other 30$~\%$ is in the different local configuration~\cite{Watanuki,Matsukawa2014}. Hence, there exists more number of variations of the local environment around Yb in the quasicrystal than in the AC. Since the quasicrystal possesses no periodicity of the lattice and hence has no unit cell, it is expected that the valence QCP spots are widespread and condensed in the ground-state phase diagram, while the number of the spots in the AC is bounded at most up to 24 since the unit cell contains the 24 Yb atoms. This can be the origin that causes quite different pressure dependences of the quantum criticality in both systems~\cite{WM2013,WM2015}: In the quasicrystal, the quantum criticality is robust under pressure at least up to $P=1.6$~GPa~\cite{Deguchi}, while in the AC, the quantum criticality appears in the vicinity of $P=1.96$~GPa~\cite{Matsukawa2016}. 

It is also noted that as a peculiar feature in the present system, the expectation value of the quadrupole operator $O_{\Gamma}$ has a non-zero value, where $\Gamma=x^2$, $y^2$, $z^2$, $xy$, $yz$, and $zx$. 
In usual high-symmetry lattices, the CEF Hamiltonian has a few finite values of the matrix elements and hence the eigenvector is expressed as the superposition of the $|J_{z}\rangle$ states with the non-zero matrix elements. Since the quadrupole operator requires the matrix elements between $|J_z\rangle$ and up to $|J_z\pm2\rangle$ states, $\langle\Psi_{\pm}|O_{\Gamma}|\Psi_{\pm}\rangle$ becomes zero for the Kramers doublet 
in the lattices with high symmetry such as cubic and tetragonal crystals. 
However, in the present system with low symmetry, most of diagonal and off-diagonal elements of $H_{\rm CEF}$ have non-zero values, which result in Eqs.~(\ref{eq:WFup}) and (\ref{eq:WFdw}) with non-zero coefficients $a_i\ne 0$. 
Then, $\langle\Psi_{\pm}|O_{\Gamma}|\Psi_{\pm}\rangle$ has non zero values for $O_{x^2}=J_{x}^2$, $O_{y^2}=J_{y}^2$, $O_{z^2}=J_{z}^2$, $O_{xy}=J_{x}J_{y}+J_{y}J_{x}$, $O_{yz}=J_{y}J_{z}+J_{z}J_{y}$, and $O_{zx}=J_{z}J_{x}+J_{x}J_{z}$. 

This offers an interesting possibility in the ultrasound measurement.  
The quadrupole susceptibility is defined as  
\begin{eqnarray}
\chi_{\Gamma}({\bf q}, \omega)&=&\sum_{\sigma_1,\sigma_2,\sigma_3,\sigma_4}
\langle\Psi_{\sigma_1}|O_{\Gamma}|\Psi_{\sigma_2}\rangle
\langle\Psi_{\sigma_3}|O_{\Gamma}|\Psi_{\sigma_4}\rangle
\nonumber
\\
& &\times
\chi_{\sigma_1\sigma_2\sigma_3\sigma_4}({\bf q}, \omega), 
\label{eq:chi_g}
\end{eqnarray}
where $\chi_{\sigma_1\sigma_2\sigma_3\sigma_4}({\bf q}, \omega)$ is the irreducible susceptibility with the momentum $\bf q$ and frequency $\omega$ for the CEF ground state $\sigma_i=+,-$. 
The elastic constant is generally expressed as 
\begin{eqnarray}
C_{\Gamma}=C_{\Gamma}^{(0)}-g_{\Gamma}^2\chi_{\Gamma}, 
\label{eq:ElasticC}
\end{eqnarray}
where $C_{\Gamma}^{(0)}$ is the elastic constant of the background including unharmonicity of the lattice   
and $g_{\Gamma}$ is the quadrupole-strain coupling constant~\cite{Thalmeier}. Here, $\chi_{\Gamma}$ is the ``$k$ limit" of the quadrupole susceptibility Eq~(\ref{eq:chi_g}), i.e., $\chi_{\Gamma}=\lim_{{\bf q}\to{\bf 0}}\lim_{\omega\to 0}\chi_{\Gamma}({\bf q},\omega)$~\cite{Kontani}. 
Since the form factors with $O_{\Gamma}$ in Eq.~(\ref{eq:chi_g}) are non-zero in the quasicrystal and AC, it is interesting to detect the softening by the ultrasound measurement. 

%%%%%
%\textcolor{red}
%{
It should be also noted that absence of the inversion symmetry allows the possibility of the finite expectation value of the odd-parity quantities such as the electronic dipole and the magnetic quadrupole, which is usually forbidden in crystals with the inversion symmetry. 
%}
%%%%%

Concluding this section, we have discussed the property of the CEF at Yb on the AC, whose local environment is common to that of 70~$\%$ of Yb in the quasicrystal. 
Although the point charge model 
%under an assumption that the valence of Yb is 3+ 
is useful as the first step analysis, the effect of the f-c hybridization can be also important for the CEF level scheme~\cite{TKasuya1985}. 
Hence, when one makes the comparison with experiments quantitatively, this point should be kept in one's mind. However, the qualitative nature of the CEF noted above is considered to be still valid even after taking account of the hybridization effect.  

%-------------------------------------------------------------------------------------
\section{Lattice Model with Onsite Interorbital Interaction and Numerical Results}

In this section, we construct the effective model for the electronic state on the AC by introducing the onsite interorbital Coulomb repulsion between the 4f- and 5d- orbitals at Yb site and discuss the ground-state property on the basis of the numerical calculation.

%-----------------------------------------------------------------------------
\subsection{Extended periodic Anderson model}

Recent measurement in the Yb-Au-Al quasicrystal with Al replaced by Ga has revealed that the quantum critical behavior in physical quantities disappears~\cite{Matsukawa2014}. This suggests that the conduction electrons on the Al sites contribute to the quantum critical state. 
Thus, as the simplest minimal model, the extended periodic Anderson model with the 4f orbital at Yb and the 3p orbital at Al has been constructed and used for the theoretical analysis, so far~\cite{WM2013,WM2015,WM2016}. 
The microscopic origin to cause the quantum valence criticality is considered to be the Coulomb repulsion between electrons on the 4f and conduction-electron orbitals~\cite{WM2010}, which has been taken into account in the previous studies~\cite{WM2013,WM2015,WM2016} as the Coulomb repulsion between electrons on the 4f and 3p orbitals in the two-orbital model. However, 
in reality, as noted in Ref.~\cite{WM2013}, the Coulomb repulsion between electrons on the 4f and 5d orbitals at the Yb site is considered to be important since it is the onsite interaction and hence is expected to give the dominant contribution to the interorbital Coulomb repulsion. 

In this paper, we analyze this effect by taking into account the onsite 4f-5d Coulomb interaction. To this end, we consider the minimal model, which consists of the 4f and 5d orbitals at Yb and the 3p orbital at Al, as follows: 

\begin{eqnarray}
H=H_{\rm f}+H_{\rm c}+H_{\rm hyb}+H_{U_{\rm fd}}.
\label{eq:EPAM}
\end{eqnarray}
As the first step of analysis, we consider the model in which the Al/Au mixed sites in Figs.~\ref{fig:YbAuAl}(b) and \ref{fig:YbAuAl}(d) are all occupied by Al and the orbital degeneracy is neglected. 
%%%%%
%\textcolor{red}
%{
As discussed in Sect.~2, the essential feature of the quasicrystal is the emergence of the condensed valence QCPs in the ground-state phase diagram because of the Al/Au mixed 
%}
%%%%%
%%%%%
%\textcolor{red}
%{
sites. This point will be discussed later in Sect.~3.4. 
%}
%%%%%
Because of the relatively small existence ratio of the 1st shell [Fig.~\ref{fig:YbAuAl}(a)], we take the shell structures from the 2nd shell to the 5th shell into account. Since we concentrate on properties in the ground state and low-temperature region much smaller than the first-excitation energy of the CEF, the hole picture is taken in Eq.~(\ref{eq:EPAM}).
 
The 4f-hole part $H_{\rm f}$ in Eq.~(\ref{eq:EPAM}), for the 4f hole on Yb, is given by
\begin{eqnarray}
H_{\rm f}=
\sum_{j=1}^{N_{\rm L}}
\left[
\varepsilon_{\rm f}\sum_{i=1\sigma}^{24}n_{ji\sigma}^{\rm f}
+U\sum_{i=1}^{24}n_{ji\uparrow}^{\rm f}n_{ji\downarrow}^{\rm f}
\right],
\label{eq:Hf}
\end{eqnarray}
where 
$\varepsilon_{\rm f}$ is the 4f-hole level, $U$ is the Coulomb repulsion between the 4f holes, and $N_{\rm L}$ is the number of the unit cells. 
Here, $n_{ji\sigma}^{\rm f}\equiv f_{ji\sigma}^{\dagger}f_{ji\sigma}$ is the number operator for the 4f hole on the $i$th site in the $j$th unit cell with ``spin" $\sigma=\uparrow, \downarrow$, which specifies the Kramers doublet of the CEF ground state, $|\Psi_{+}\rangle$ and $|\Psi_{-}\rangle$ in Eqs.~(\ref{eq:WFup}) and (\ref{eq:WFdw}), respectively. 
%%%%%
%\textcolor{red}
%{
Here we consider the case where $\varepsilon_{\rm f}$ is the same for all the Yb sites and the effect of the distribution of $\varepsilon_{\rm f}$ due to the Al/Au mixed sites will be discussed in Sect. 3.4, as noted above. 
%}
%%%%%

The conduction-hole part $H_{\rm c}$ in Eq.~(\ref{eq:EPAM}) is given by 
\begin{eqnarray}
H_{\rm c}&=&
\sum_{\langle j{\xi}, j'{\nu}\rangle\sigma}
\left(t_{j{\xi}, j'{\nu}}^{\rm c}
c_{j\xi\sigma}^{\dagger}c_{j'\nu\sigma}+{\rm h.c.}
\right)
\nonumber
\\
&+&\sum_{\langle ji, j'i'\rangle\sigma}
\left(t_{ji, j'i'}^{\rm d}
d_{ji\sigma}^{\dagger}d_{j'i'\sigma}+{\rm h.c.}
\right)
\nonumber
\\
&+&\sum_{\langle ji, j'\xi\rangle\sigma}
\left(V_{ji, j'\xi}^{\rm d}
d_{ji\sigma}^{\dagger}c_{j'\xi\sigma}+{\rm h.c.}
\right), 
\label{eq:Hc}
\end{eqnarray}
where the first, second, and third lines are the transfer for the 3p holes on Al, 
that for the 5d holes on Yb, and the hybridization between the 3p and 5d holes, respectively. 
Here, $\langle j{\xi}, j'{\nu}\rangle$ denotes the pairs between the $j\xi$th Al site and the $j'\nu$th Al site. $\langle ji, j'i'\rangle$ denotes the pairs between the $ji$th Yb site and the $j'i'$th Yb site. 
$\langle ji, j'\xi\rangle$ denotes the pairs between the $ji$th Yb site and the $j'\xi$th Al site.
To parametrize transfer integrals,  
we follow the argument of the linear combination of atomic orbitals (LCAO) and 
employ the relation $t_{ji,j'i'}^{\rm d}\propto 1/r^{\ell+\ell'+1}$, where 
$r$ is the distance between the center of the 5d orbitals with azimuthal quantum numbers, $\ell=2$ and $\ell'=2$, respectively~\cite{Andersen1,Andersen2}.
As for the transfers for p orbitals, we set $t_{j\xi,j'\nu}^{\rm c}\propto r^{-2}$, following the Harrison's argument for free electrons~\cite{Harrison}. 
%%%%%
%\textcolor{red}
%{
Here, the energy level of the Al-3p state is set to be the origin of the energy. The Yb-5d level is set to be $0$ for simplicity since the main result below is expected to be unchanged as far as the 5d band has a certain filling.  
%}
%%%%%

The hybridization between the 4f hole on Yb and the 3p hole on Al, $H_{\rm hyb}$ in Eq.~(\ref{eq:EPAM}), is given by 
\begin{eqnarray}
H_{\rm hyb}=
\sum_{\langle ji, j'\xi\rangle\sigma}
\left(V_{ji, j'\xi}^{\rm f}
f_{ji\sigma}^{\dagger}c_{j'\xi\sigma}+{\rm h.c.}
\right).    
\label{eq:Hhyb}
\end{eqnarray}
To parametrize hybridizations, 
we employ the relation $V_{ji,j'\xi}^{\rm f(d)}\propto 1/r^{\ell+\ell'+1}$, where $r$ is the distance between the 4f (5d) orbital and the 3p orbital with azimuthal quantum numbers, $\ell=3 (2)$ and $\ell'=1$, respectively~\cite{Andersen1,Andersen2}. 

The inter-orbital Coulomb repulsion between the 4f hole and 5d hole on Yb, $H_{U{\rm fd}}$ in Eq.~(\ref{eq:EPAM}),  is given by
\begin{eqnarray}
H_{U_{\rm fd}}=U_{\rm fd}\sum_{j=1}^{N_{\rm L}}\sum_{i=1}^{24}n_{ji}^{\rm f}n_{ji}^{\rm d},  
\label{eq:HUfc}
\end{eqnarray}
where 
$n_{ji\sigma}^{\rm d}\equiv d_{ji\sigma}^{\dagger}d_{ji\sigma}$ is the number operator for the 5d hole on the $ji$th site with ``spin" $\sigma$. 

The largest interaction in the present model is the on-site Coulomb repulsion $U$ in Eq.~(\ref{eq:Hf}), which is considered to be the origin of the emergence of the heavy electron state $\lim_{T\to 0}C(T)/T=\gamma\sim 700$~mJ/K$^2$ mol-Yb in the AC at ambient pressure. 
To analyze the heavy electron state, we apply the slave-boson mean-field theory~\cite{Read,OM2000} to Eq.~(\ref{eq:EPAM}). 
To describe the state with $U=\infty$, we consider $V_{ji, j'\xi}f_{ji\sigma}^{\dagger}b_{i}c_{j'\xi\sigma}$ instead of $V_{ji, j'\xi}f_{ji\sigma}^{\dagger}c_{j'\xi\sigma}$ in Eq.~(\ref{eq:Hhyb}) 
by introducing the slave-boson operator $b_i$ at the $i$th site 
in the $j$th unit cell to describe the $f^0$ state 
and require the constraint $\sum_{\sigma}n^{\rm f}_{ji\sigma}+b^{\dagger}_{i}b_{i}=1$ with introducing the Lagrange multiplier $\lambda_{i}$, i.e., $\sum_{i=1}^{24}\lambda_{i}(\sum_{\sigma}n^{\rm f}_{ji\sigma}+b^{\dagger}_{i}b_{i}-1)$. 
Hereafter, we treat $b_i$ as the mean field: $\overline{b_i}=\langle b_{i}\rangle$.

As for $H_{U_{\rm fd}}$, we treat in the mean-field approximation: 
\begin{eqnarray}
U_{\rm fd}\sum_{j=1}^{N_{\rm L}}\sum_{i=1}^{24}n_{ji}^{\rm f}n_{ji}^{\rm d}&\approx&\sum_{\bf k}\sum_{i=1}^{24}\left[ U_{\rm fd}
%%%%%
%\textcolor{red}
%{ 
\langle n_{i}^{\rm f}
%}
%%%%%
\rangle n_{{\bf k}i}^{\rm d}+R_{i}n_{{\bf k}i}^{\rm f}\right]
\nonumber
\\
& &
-N_{\rm L}\sum_{i=1}^{24}R_{i}\langle n_{i}^{\rm f}\rangle, 
\end{eqnarray}
where the mean-field $R_i$ is defined by 
%%%%%
%\textcolor{red}
%{
\begin{eqnarray}
R_i\equiv U_{\rm fd}\langle n_{i}^{\rm d}\rangle. 
\label{eq:Rdef}
\end{eqnarray}
%}
%%%%%
Then, the resultant mean-field Hamiltonian $\tilde{H}$ is expressed as 
\begin{eqnarray} 
\tilde{H}=\sum_{{\bf k}\sigma}\tilde{H}_{{\bf k}\sigma}
+N_{\rm L}\sum_{i=1}^{24}\lambda_{i}(\overline{b_{i}}^2-1)-N_{\rm L}\sum_{i=1}^{24}R_{i}\langle n_{i}^{\rm f}\rangle, 
\label{eq:H_MF_full}
\end{eqnarray}
where $\tilde{H}_{{\bf k}\sigma}$ is given by 
\begin{eqnarray}
\tilde{H}_{{\bf k}\sigma}&=&
\sum_{i=1}^{24}
(\varepsilon_{\rm f}+\lambda_{i}+R_{i})
f_{{\bf k}i\sigma}^{\dagger}f_{{\bf k}i\sigma}
\nonumber
\\
&+&\sum_{\langle\xi\nu\rangle}
\left(
t_{{\bf k}\xi\nu}^{\rm c}c_{{\bf k}\xi\sigma}^{\dagger}c_{{\bf k}\nu\sigma}+{\rm h.c.}
\right)
\nonumber
\\
&+&
\sum_{i=1}^{24}
U_{\rm fd}\langle n_{i}^{\rm f}\rangle
d_{{\bf k}i\sigma}^{\dagger}d_{{\bf k}i\sigma}
+\sum_{\langle i i'\rangle}
\left(
t_{{\bf k}ii'}^{\rm d}d_{{\bf k}i\sigma}^{\dagger}d_{{\bf k}i'\sigma}+{\rm h.c.}
\right)
\nonumber
\\
&+&\sum_{\langle{i}\xi\rangle}
\left(
V_{{\bf k}i\xi}^{\rm d}d_{{\bf k}i\sigma}^{\dagger}
c_{{\bf k}\xi\sigma}+{\rm h.c.}
\right)
\nonumber
\\
&+&\sum_{\langle{i}\xi\rangle}
\left(
V_{{\bf k}i\xi}^{\rm f}f_{{\bf k}i\sigma}^{\dagger}
\overline{b_{i}}c_{{\bf k}\xi\sigma}+{\rm h.c.}
\right). 
\label{eq:HMF}
\end{eqnarray}

By optimizing the ground-state energy with respect to $\lambda_i$, $\overline{b_i}$, and $R_i$, $\partial\langle \tilde{H}\rangle/\partial\lambda_i=0$, $\partial\langle \tilde{H}\rangle/\partial\overline{b_i}=0$, and $\partial\langle \tilde{H}\rangle/\partial R_i=0$, we obtain the mean-field equations as
\begin{eqnarray}
\frac{1}{N_{\rm L}}\sum_{{\bf k}\sigma}
\langle f_{{\bf k}i\sigma}^{\dagger}f_{{\bf k}i\sigma}\rangle
+\overline{b_{i}}^2&=&1, 
\label{eq:MF1}
\\
\frac{1}{2N_{\rm L}}\sum_{{\bf k}\sigma}
\sum_{\xi}
\left[
V_{{\bf k}i\xi}
\langle f^{\dagger}_{{\bf k}i\sigma}c_{{\bf k}\xi\sigma}
\rangle
+{\rm h.c.}
\right]
+\lambda_{i}\overline{b_{i}}&=&0, 
\label{eq:MF2}
\\
%R_i-U_{\rm fd}\langle n_{i}^{\rm d}\rangle&=&0, 
%%%%%
%\textcolor{red}
%{
\langle n_{i}^{\rm f}\rangle =\frac{1}{N_{\rm L}}\sum_{\bf k}\langle n_{{\bf k}i}^{\rm f}\rangle, 
%}
%%%%%
\label{eq:MF3}
\end{eqnarray}
respectively, for $i=1, .., 24$.
%%%%%
%\textcolor{red}
%{ 
Since Eq. (\ref{eq:MF3}) is satisfied by definition, 
we solve Eqs.~(\ref{eq:Rdef}), (\ref{eq:MF1}), and (\ref{eq:MF2})   
%}
%%%%%
and 
the equation for the filling, $\bar{n}=1$, self-consistently. 
Here, the filling $\bar{n}$ is defined by the hole number per site, which is given by 
$\bar{n}=\bar{n}_{\rm f}+\bar{n}_{\rm c}+\bar{n}_{\rm d}$
, where $\bar{n}_{\rm f}$, $\bar{n}_{\rm c}$, and $\bar{n}_{\rm d}$ 
are defined as 
$\bar{n}_{\rm f}\equiv\frac{1}{N_{\rm L}}\sum_{j=1}^{N_{\rm L}}\frac{1}{24}\sum_{i=1}^{24}\sum_{\sigma}\langle n_{ji\sigma}^{\rm f}\rangle$, 
$\bar{n}_{\rm c}\equiv\frac{1}{N_{\rm L}}\sum_{j=1}^{N_{\rm L}}
\frac{1}{48}\sum_{\xi=1}^{48}\sum_{\sigma}\langle n_{j\xi\sigma}^{\rm c}\rangle$ 
with $n_{j\xi\sigma}^{\rm c}\equiv c_{j\xi\sigma}^{\dagger}c_{j\xi\sigma}$, and 
$\bar{n}_{\rm d}\equiv\frac{1}{N_{\rm L}}\sum_{j=1}^{N_{\rm L}}\frac{1}{24}\sum_{i=1}^{24}\sum_{\sigma}\langle n_{ji\sigma}^{\rm d}\rangle$, respectively. 

To estimate the parameters in Eq.~(\ref{eq:EPAM}), we follow Harrison~\cite{Harrison}: 
The transfer integral for p orbitals between the nearest-neighbor (N.N.) Al sites on the 2nd shell (Fig.~\ref{fig:YbAuAl}(b)) is estimated as $t_2^{\rm p}\approx 4.0$~eV by inputting the Al-Al distance $r=2.48$~\AA \ to $t_2^{\rm p}=\eta_{{\rm pp}\sigma}\frac{\hbar^2}{m_e}\frac{1}{r^2}$, where $\eta_{{\rm pp}\sigma}=3.24$, $\hbar$ is the reduced Planck constant, and $m_e$ is the bare electron mass~\cite{Harrison}. 
In the similar way, the transfer integral for d orbitals between the N.N. Yb sites on the 3rd shell (Fig.~\ref{fig:YbAuAl}(c)) is estimated as $t^{\rm d}_{3}\approx-0.03$~eV by inputting the Yb-Yb distance $r=5.44$~\AA. 
The hybridization for the 5d orbital and 3p orbital between the N.N. Yb and Al sites on the 3rd and 2nd shells, respectively, is estimated as $V^{\rm d}_{32}\approx -0.42$~eV by inputting the Yb-Al distance $r=3.11$~\AA. 

These values are expressed in the unit of $t_2^{\rm p}$ precisely as 
$t_3^{\rm d}/t_2^{\rm p}=-0.0064$ and 
$V_{32}^{\rm d}/t_2^{\rm p}=-0.1057$. 
Here we set the hybridization between the 4f orbital and 3p orbital at the N.N. Yb and Al sites on the 3rd and 2nd shells as $V_{\rm 32}^{\rm f}/t_2^{\rm p}=-0.1$, whose absolute value is slightly smaller than $|V_{32}^{\rm d}/t_2^{\rm p}|$ as a reasonable value. In the following calculation, we use these values as input parameters in Eq.~(\ref{eq:EPAM}) and the other parameters such as $t_{j\xi,j'\nu}^{\rm p}$, $t_{ji,j'i'}^{\rm d}$, $V_{ji,j'\xi}^{\rm d}$, and $V_{ji,j'\xi}^{\rm f}$ are set so as to follow the distance dependence as mentioned above. 
Hereafter, the energy unit is taken as $t_2^{\rm p}=1$. 
The calculation is performed in $N_{\rm L}=8\times 8\times 8$.

%-----------------------------------------------------------------------------
\subsection{The f-level dependence of the 4f-hole number}

\begin{figure}[h]
\includegraphics[width=7cm]{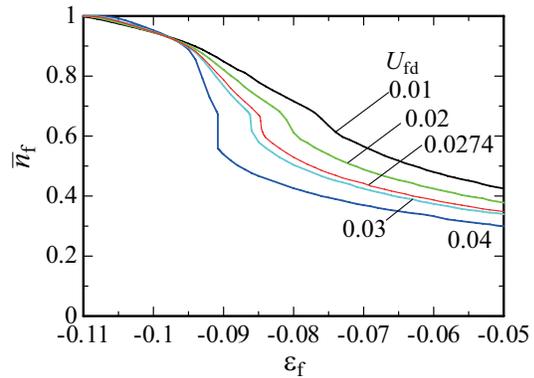}
\caption{(Color online) The $\varepsilon_{\rm f}$ dependence of $\bar{n}_{\rm f}$ for $U_{\rm fd}=0.01$ (black), $0.02$ (green), $0.0274$ (red), $0.03$ (light blue), and $0.04$ (blue).} 
\label{fig:nf_Ef}
\end{figure}

In this subsection, we discuss the result on the $\varepsilon_{\rm f}$ dependence of the 4f-hole number $\bar{n}_{\rm f}$, as shown in Fig.~\ref{fig:nf_Ef}. 
Note that $\bar{n}_{\rm f}=1$ $(\bar{n}_{\rm f}=0)$ corresponds to Yb$^{+3}$ (Yb$^{+2}$). 
For $U_{\rm fd}=0.01$, the 4f-hole number $\bar{n}_{\rm f}$ gradually increases as $\varepsilon_{\rm f}$ decreases.
As $U_{\rm fd}$ increases, $\bar{n}_{\rm f}$ tends to increase sharply around $\varepsilon_{\rm f}\sim -0.08$. 
For $U_{\rm fd}=0.0274$, the slope $\partial \bar{n}_{\rm f}/\partial\varepsilon_{\rm f}$ diverges at $\varepsilon_{\rm f}=-0.0849$. As $U_{\rm fd}$ further increases, a jump $\Delta n_{\rm f}$ in ${\bar n}_{\rm f}$ starts to appear: i.e., $\Delta n_{\rm f}=0.052$ for $U_{\rm fd}=0.03$ and $\Delta n_{\rm f}=0.113$ for $U_{\rm fd}=0.04$. 
The emergence of the jump in $\bar{n}_{\rm f}$ indicates that the first-order valence transition (FOVT) takes place.

%%%%%%%%%%%%  Fig.3  %%%%%%%%%%%%%%%%%%%%%%%%%%%%%%
%%%%%%%%%%%%%%%%%%%%%%%%%%%%%%%%%%%%%%%%%%%%%%%%%%%

\begin{figure}[h]
\includegraphics[width=7cm]{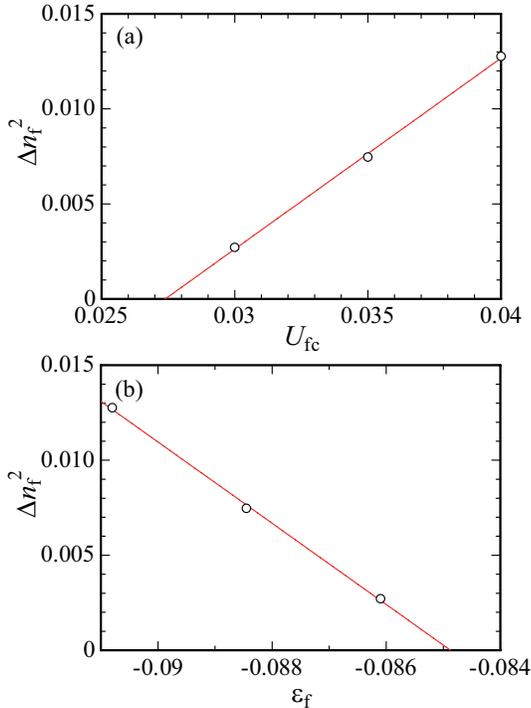}
\caption{(Color online) (a) $\Delta n_{\rm f}^2$ vs. $U_{\rm fd}$. (b) $\Delta n_{\rm f}^2$ vs. $\varepsilon_{\rm f}$.} 
\label{fig:dnf_Ufc}
\end{figure}

In order to determine the location of the QCP precisely, we plot the $U_{\rm fd}$ dependence of $\Delta n_{\rm f}^2$ in Fig.~\ref{fig:dnf_Ufc}(a). 
As the result obtained by the mean-field theory is expected to give $\Delta n_{\rm f}=a(U_{\rm fd}-U_{\rm fd}^{\rm QCP})^{\beta}$ with the critical exponent $\beta=1/2$, the data follow a straight line. 
The least-square fit of the data actually gives $\beta=0.5$ and the QCP is identified as the point where $\Delta n_{\rm f}^2\to 0$, that is, $U_{\rm fd}^{\rm QCP}=0.0274$. 

In the same way, we plot the $\varepsilon_{\rm f}$ dependence of $\Delta n_{\rm f}^2$ in Fig.~\ref{fig:dnf_Ufc}(b). The least-square fit of the data gives $\Delta n_{\rm f}=b(\varepsilon_{\rm f}^{\rm QCP}-\varepsilon_{\rm f})^{\beta}$ with $\beta=0.5$ and the QCP is identified as the point where $\Delta n_{\rm f}^2\to 0$ to be $\varepsilon_{\rm f}^{\rm QCP}=-0.0849$. 

%%%%%%%%%%%%  Fig.5  %%%%%%%%%%%%%%%%%%%%%%%%%%%%%%
%%%%%%%%%%%%%%%%%%%%%%%%%%%%%%%%%%%%%%%%%%%%%%%%%%%

\begin{figure}[h]
\includegraphics[width=7cm]{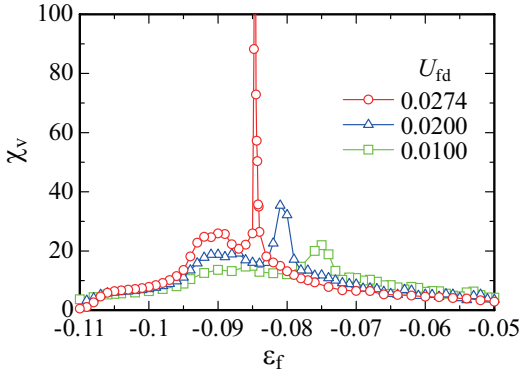}
\caption{(Color online) The valence susceptibility $\chi_{\rm v}\equiv-\partial \bar{n}_{\rm f}/\partial\varepsilon_{\rm f}$ for $U_{\rm fc}=0.01$ (square), $0.02$ (triangle), and 0.0274 (circle).} 
\label{fig:chi_v}
\end{figure}

Thus, the QCP is identified as $(\varepsilon_{\rm f}^{\rm QCP}, U_{\rm fd}^{\rm QCP})=(-0.0849, 0.0274)$. 
The value of $\bar{n}_{\rm f}$ at the QCP is $n_{\rm f}=0.67$. Hence, our model [Eq.~(\ref{eq:EPAM})] has shown that the valence QCP is realized with the intermediate valence of Yb, Yb$^{+2.67}$, for realistic parameters set. 
As noted above, the slope of $\bar{n}_{\rm f}$ for $U_{\rm fd}=0.0274$ diverges at $\varepsilon_{\rm f}=-0.0849$ in Fig.~\ref{fig:nf_Ef}, which indicates that the valence susceptibility diverges: $\chi_{\rm v}\equiv-\partial \bar{n}_{\rm f}/\partial\varepsilon_{\rm f}=\infty$ (see Fig.~\ref{fig:chi_v}). 
Namely, the CVF diverges at the QCP. 
Since $U_{\rm fd}$ is the on-site 4f-5d interaction at Yb, the value of 
$U_{\rm fd}^{\rm QCP}=0.0274t_{2}^{\rm p}=0.1096$~eV is considered to be possible to be realized in the actual material in the following reason.

The location of the valence QCP was studied in the extended periodic Anderson model which consists of the 4f and conduction electrons in the one-spatial dimension (1d)~\cite{WM2006} and infinite dimension~\cite{Saiga2008}. In the 1d system, by the slave-boson mean-field calculation for the f-c hybridization $V=0.1$, whose absolute value is the same as the present case $|V_{32}^{\rm f}|=0.1$,  the QCP was identified as $U_{\rm fc}^{\rm QCP}=0.98$~\cite{WM2006}. However, by the density matrix renormalization group (DMRG) calculation in the same model, it was shown that the valence-crossover (VCO) region is strongly stabilized by the effects of quantum fluctuation and electron correlation and the QCP was identified as $U^{\rm QCP}_{\rm fc}=5.9$~\cite{WM2006}. Namely, the correct value is about 6-times larger than that by the mean-field calculation. Although this value may be overestimated for real materials because of too-strong quantum fluctuation specific to 1d, if we estimate the correct value of $U^{\rm QCP}_{\rm fd}$ in the present system by multiplying the factor 6 to the mean-field result, we obtain $U^{\rm QCP}_{\rm fd}\approx 0.66$~eV.

  In this paper, we employ the LCAO argument~\cite{Andersen1,Andersen2} to evaluate the parameters in Eq.~(\ref{eq:EPAM}). For more precise evaluation, the constrained RPA method for the Wannier orbitals whose energy band is located at the Fermi level $E_{\rm F}$ constructed from the downfolding procedure is promising~\cite{cRPA}. In the procedure, if the 5d orbital at Yb contributes dominantly to the Wannier orbital for conduction electrons whose energy band is located at $E_{\rm F}$, the resultant $U_{\rm fd}$ obtained by performing the spatial integration of the Coulomb repulsion between the Wannier orbitals for the 4f and conduction electrons is expected to have a large value. 

The Ce metal, which exhibits the FOVT well-known as the $\gamma$-$\alpha$ transition, is considered to be the case because the energy bands located at $E_{\rm F}$ mainly consists of the 4f and 5d orbitals at Ce~\cite{Pickett}. The onsite Coulomb repulsion $U$ between the 4f electrons in the Ce metal was evaluated by the constrained RPA method, which is $U\approx 3\sim 8$~eV depending on the choice of the energy window for the targeted low-energy region~\cite{Ce_Uff}. In any case, $U_{\rm fd}$ is expected to be smaller than this value since it is the interorbital interaction. 

  In the AC Yb$_{14}$Al$_{35}$Au$_{51}$, the 3p electron at Al is considered to contribute to the energy band located at $E_{\rm F}$ as mentioned above so that the rate of the 5d orbital at Yb contributing to the Wannier orbital for the conduction electron may not be so large. In that case, $U_{\rm fd}$ is expected to be smaller than that in the Ce metal. From these considerations, the value we estimated above, $U^{\rm QCP}_{\rm fd}\approx 0.66$~eV seems to be a reasonable value although it should be examined quantitatively from the first principles in the future.

%%%%%%%%%%%%  Fig.6  %%%%%%%%%%%%%%%%%%%%%%%%%%%%%%
%%%%%%%%%%%%%%%%%%%%%%%%%%%%%%%%%%%%%%%%%%%%%%%%%%%

%-----------------------------------------------------------------------------
\subsection{The ground-state phase diagram}

\begin{figure}[h]
\includegraphics[width=7cm]{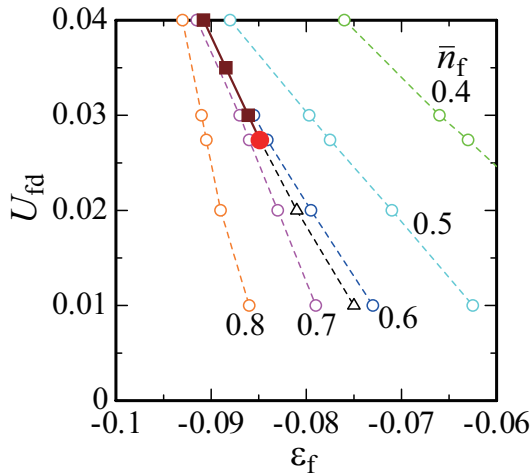}
\caption{(Color online) Ground state phase diagram in $U_{\rm fd}$-$\varepsilon_{\rm f}$ plane. The first-order valence transition (solid line with filled square) terminates at the QCP (filled circle). The valence crossover line (dashed line with open triangle) extends from the QCP.
%%%%%
%\textcolor{blue}
%{
The dashed lines with open circles are contour lines for $\bar{n}_{\rm f}=0.4, 0.5, 0.6, 0.7$, and $0.8$. 
%}
%%%%%
} 
\label{fig:PD}
\end{figure}

In this subsection, we discuss the ground-state phase diagram in the $\varepsilon_{\rm f}$-$U_{\rm fd}$ plane, as shown in Fig.~\ref{fig:PD}. The FOVT line (solid line
%%%%%
%\textcolor{blue}
%{
with squares) 
%} 
%%%%%
terminates at the QCP. The VCO line (dashed line
%%%%%
%\textcolor{blue}
%{
with triangles) 
%}
%%%%%
 where the CVF develops with enhanced $\chi_{\rm v}$ (see Fig.~\ref{fig:chi_v}) extends from the QCP. 
%%%%%
%\textcolor{blue}
%{
We also plot the contour lines for $\bar{n}_{\rm f}=0.4$, 0.5, 0.6, 0.7, and 0.8 as dashed lines with open circles. 
%}
%%%%%
In the deeper-$\varepsilon_{\rm f}$ side, the relatively larger-$\bar{n}_{\rm f}$ state is realized, continuing to the Kondo state with $n_{\rm f}=1$ realized in the deep-$\varepsilon_{\rm f}$ regime. 
In the side with shallower $\varepsilon_{\rm f}$, the relatively smaller-$\bar{n}_{\rm f}$ state is realized, which is called the mixed-valence regime. 
Both the states have the same symmetry with the large Fermi surface, i.e., $\langle f^{\dagger}_{{\bf k}i\sigma}c_{{\bf k}\xi\sigma}\rangle$ is finite at everywhere in the phase diagram, which can be continuously connected by circumventing the QCP via the crossover with the Luttinger's sum rule being satisfied.

%%%%%%%%%%%%  Fig.7  %%%%%%%%%%%%%%%%%%%%%%%%%%%%%%
%%%%%%%%%%%%%%%%%%%%%%%%%%%%%%%%%%%%%%%%%%%%%%%%%%%

%%%%%
%\textcolor{blue}
%{
It is noted that we performed the calculation for $V_{32}^{\rm f}=-0.2$ and the similar analysis as in Figs.~\ref{fig:dnf_Ufc}(a) and \ref{fig:dnf_Ufc}(b) gives $(\varepsilon_{\rm f}^{\rm QCP}, U_{\rm fd}^{\rm QCP})=(-0.2217, 0.0693)$. Namely, as the magnitude of the f-c hybridization increases, the location of the valence QCP shifts to the left-upper direction in Fig.~\ref{fig:PD}. This point will be discussed in Sect.~3.4 in relation to the lattice-constant dependence of the Yb valence. 
%}
%%%%%

In the present calculation, we restrict ourselves within the paramagnetic state. However, it is also possible to discuss a possible existence of magnetic state. Such a calculation was performed in the extended periodic Anderson model on the square lattice as a model of CeRhIn$_5$ under the pressure and the magnetic field~\cite{WM2010}. The result was that for the large f-c hybridization $V$, the magnetically-ordered phase appears in the deep-$\varepsilon_{\rm f}$ regime, which is far separated from the FOVT and VCO line in Fig.~\ref{fig:PD}. Then, the magnetic transition point (i.e., the magnetic QCP) and the FOVT point or VCO point are separated in this case~\cite{WM2011} because the magnetic state is suppressed by growing the f-c hybridization. 
On the other hand, as $V$ decreases, the magnetically-ordered phase extends toward the shallower-$\varepsilon_{\rm f}$ region and finally reach the FOVT and VCO line. Even in the much smaller region of $V$, the coincidence is kept because the sudden jump in $\bar{n}_{\rm f}$ at the FOVT line and the enhanced CVF at the VCO line suppresses the magnetic order, which makes the concurrence retained. 
Interestingly, the slave-boson mean-field calculation showed that the resultant magnetic transition becomes the first-order transition even at the VCO line~\cite{WMCeRhIn5}.  

In the case of the AC, the N.N. distance between the Yb site and Al site is 3.07~\AA \ so that  the f-c hybridization is rather small as we set $|V_{32}^{\rm fp}|=0.1/t_{2}=0.1$ in the end of Sect.~3.1. Hence, if we take into account the magnetic state in the present calculation, it is expected that the magnetically-ordered phase appears in just the left-hand side of the FOVT line in Fig.~\ref{fig:PD} and the coincidence is realized.  
Experimentally, the magnetic order has been observed in the AC~\cite{Matsukawa2016} just after the emergence of the divergent behavior in $\chi$ at $P=1.96$~GPa as the pressure increases, which is considered to be due to the CVFs. 
The coincidence of the magnetic transition point and the sharp VCO point is expected to occur generally, which has been reported recently by the transport measurement in CeRhIn$_5$~\cite{Ren2017}. 

%-------------------------------------------------------------------------
\subsection{Lattice constant dependence of the Yb valence}

Recently, the systematic variations of the elements of Al and Au [e.g., Ga (Cu) is replaced by Al (Au)] in the Yb-Al-Au quasicrystal and AC have been synthesized by Nanba {\it et al}.~\cite{Nanba2017}. Interestingly, 
they plotted the Yb valence measured at $T=300$~K as a function of the six-dimensional lattice constants of the quasicrystals $a_{6{\rm D}}$~\cite{QC_text} and the lattice constants of the ACs $a$ (see discussion below), and 
discovered a tendency that all the data are on a single line. 
A remarkable finding is that in the plot, 
the quasicrystal Yb$_{15}$Al$_{34}$Au$_{51}$ is just located at the point from which the Yb valence starts to change sharply. This indicates that the quasicrystal Yb$_{15}$Al$_{34}$Au$_{51}$ is just located at the valence QCP. 

Since our theory takes into account the distance dependence of the model parameters, we can perform the theoretical analysis. As for the f level, we speculate that 
the effect of the f-c hybridization is important for the actual CEF levels in the following reason. 

When we apply the pressure to the Yb-Al-Au quasicrystal and AC, the Yb valence increases~\cite{Watanuki_Yb}. This implies that the effect of the decrease in the f-hole level 
$\varepsilon_{\rm f}$ overcomes the effect of the increase in the f-c hybridization $|V_{ji,j'\xi}^{\rm f}|$ under pressure. 
In the point charge model, the CEF ground-state energy depends on the CEF parameter $B_{\ell}^{m}$, which is the coefficient of the Stevens operator as discussed below Eq.~(\ref{eq:Vcry})~\cite{Hutchings}. Since $B_{\ell}^{m}$ has the distance dependence as $B_{\ell}^{m}\propto R_{k}^{-\ell-1}$, $B_{\ell}^{m}$ with smaller $\ell$ usually has a larger value. Actually, the recent measurement of the CEF level in the AC TbCd$_6$ has revealed that $B_{2}^{0}$ is dominant~\cite{Jazbec2016,Das2017}.  
If this contributes dominantly to the CEF in the quasicrystal Yb$_{15}$Al$_{34}$Au$_{51}$ and AC, the f level $\varepsilon_{\rm f}$ is expected to behave as $\varepsilon_{\rm f}\propto 1/r^{3}$. 
However, this $r$ dependence in $\varepsilon_{\rm f}$ cannot overcome the effect of the f-c hybridization under pressure, since the hybridization has the distance dependence as $V_{ji,j'\xi}^{\rm f}\propto 1/r^{5}$ as noted below Eq.~(\ref{eq:Hhyb}). 

On the other hand, the CEF level can arise from the f-c hybridization, as noted in Sect.~2.   
The CEF level can be evaluated by the second-order perturbation theory with respect to the f-c hybridization~\cite{TKasuya1985} and this leads to the distance dependence as  $\varepsilon_{\rm f}\propto |V_{ji,j'\xi}^{\rm f}|^2/\Delta\approx 1/r^{10}$. Here, $\Delta$ is the excitation energy to the intermediate state with the f$^0$ or f$^2$ configuration in the hole picture. 
Although the actual CEF level is considered to be contributed from both effects by the point charge model and f-c hybridization, here we proceed to our analysis as the latter effect is dominant for simplicity. 

%%%%%%%%%%%%  Fig.8  %%%%%%%%%%%%%%%%%%%%%%%%%%%%%%
\begin{figure}[h]
\includegraphics[width=7cm]{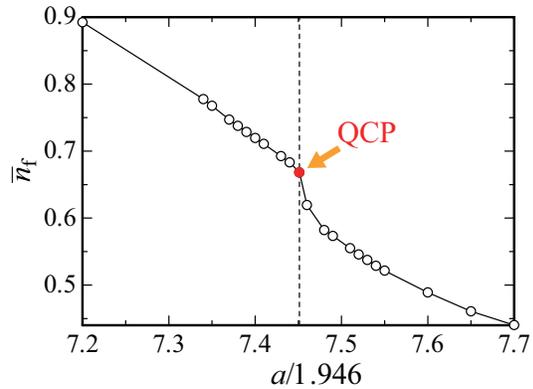}
\caption{(Color online) The 4f-hole number $\bar{n}_{\rm f}$ vs. the lattice constant $a/1.946$~\AA.
$\bar{n}_{\rm f}$ at the QCP indicated by an arrow (filled circle) is plotted at $a=14.5$~\AA \ (dashed line) (see text).} 
\label{fig:nf_a}
\end{figure}
%%%%%%%%%%%%%%%%%%%%%%%%%%%%%%%%%%%%%%%%%%%%%%%%%%%

To see the general feature of the lattice constant dependence of the Yb valence in the Yb-Al-Au quasicrystal and AC, 
in Fig.~\ref{fig:nf_a}, we plot $\bar{n}_{\rm f}$ vs. the lattice constant of the AC divided by 1.946, which corresponds to $a_{6{\rm D}}$~\cite{QC_text}. 
First, we calculate $\bar{n}_{\rm f}$ at the QCP (see Fig.~\ref{fig:PD}) indicated by an arrow at $14.5$~\AA$/1.946=7.4512$~\AA \ (vertical dashed line) in Fig.~\ref{fig:nf_a}, where $a=14.5$~\AA \ is the lattice constant of the AC at ambient pressure~\cite{Ishimasa}. 
Then, we calculate $\bar{n}_{\rm f}$ by varying the lattice constants. Here, the changes in the model parameters in Eq.~(\ref{eq:EPAM}) are all taken into account in the calculation by the distance dependence noted in Sect.~3.1 and the f level is assumed to follow $\varepsilon_{\rm f}\propto 1/r^{10}$. Here, the interorbital Coulomb repulsion $U_{\rm fd}$ is set to be the value of the QCP, $U_{\rm fd}=U_{\rm fd}^{\rm QCP}$, since it is onsite interaction at the Yb site, which is expected not to be altered severely by chemical substitution to Al and/or Au sites in the quasicrystal and AC. 

The result shows that as $a$ increases from the value at the QCP, $\bar{n}_{\rm f}$ first decreases steeply and then turns into a gradual decrease. On the other hand, as $a$ decreases from the value at the QCP, $\bar{n}_{\rm f}$ increases gradually. 
The asymmetry of the $a$ dependence of $\bar{n}_{\rm f}$ around the QCP is ascribed to the fact that increasing and decreasing lattice constants are reflected in the changes in the model parameters differently.  
When $a$ increases, the f-hole level $\varepsilon_{\rm f}$ increases and the f-c hybridization $|V_{ji,j'\xi}^{\rm f}|$ 
%%%%%
%also increases, 
%\textcolor{blue}
%{
decreases. 
Here, $U_{\rm fd}$ is considered to be unchanged because it is the onsite interaction. 
The decrease of the f-c hybridization 
makes the location of the QCP shift to the $\varepsilon_{\rm f}$-increasing and $U_{\rm fd}$-decreasing direction, i.e., the FOVT line extends in Fig.~\ref{fig:PD}. 
Hence, as $a$ increases, the increase in $\varepsilon_{\rm f}$ and the shift of the QCP occur  simultaneously. 
Since the contour lines for $\bar{n}_{\rm f}$ accompanied by the QCP also shift, 
this implies that the system with expanding $a$ proceeds to the right-upper direction from the QCP in Fig.~\ref{fig:PD}. Since this makes the system go across the relatively-dense contour area than the case where the system proceeds to the left-lower direction from the QCP in Fig.~\ref{fig:PD}, $\bar{n}_{\rm f}$ changes sharply for a certain change in $a$ (e.g., see a steep decrease in $\bar{n}_{\rm f}$ from the QCP to $a/1.946 \approx 7.475$ in Fig.~\ref{fig:nf_a}). 
%both of which contribute to make the electronic state itinerant. 
Hence, the steep decrease in $\bar{n}_{\rm f}$ immediately appears for a slight increase in $a$ from the vertical dashed line in Fig.~\ref{fig:nf_a}. 

On the other hand, when $a$ decreases, $\varepsilon_{\rm f}$ decreases while $|V_{ji,j'\xi}^{\rm f}|$ increases. 
%%%%%
%\textcolor{blue}
%{
The increase in $|V_{ji,j'\xi}^{\rm f}|$ makes the location of the QCP shift to the $\varepsilon_{\rm f}$-decreasing and $U_{\rm fd}$-increasing direction, i.e., the FOVT line shortens in Fig.~\ref{fig:PD}. 
Hence, as $a$ decreases, the decrease in $\varepsilon_{\rm f}$ and the shift of the QCP occur  simultaneously. 
Since the contour lines for $\bar{n}_{\rm f}$ accompanied by the QCP also shift, 
this implies that the system with shrinking $a$ proceeds to the left-lower direction from the QCP in Fig.~\ref{fig:PD}. Since this makes the system proceed toward the relatively-sparse contour area (i.e., the interval of the contour lines becomes spread in the left-lower direction in Fig.~\ref{fig:PD}), $\bar{n}_{\rm f}$ does not change sharply for a certain change in $a$ (e.g., see a gradual increase in $\bar{n}_{\rm f}$ from the QCP to $a/1.946\approx 7.425$ in Fig.~\ref{fig:nf_a}). 
%}
%%%%%
%Hence, both effects counteract each other and $\bar{n}_{\rm f}$ merely shows a gradual increase reflecting the decrease in $\varepsilon_{\rm f}$ as the eventual effect.
Namely, the emergence of the steep decrease in $\bar{n}_{\rm f}$ just next to the QCP reflects the enhanced CVF $\chi_{\rm v}$ as seen as the divergent slope $-\partial n_{\rm f}/\partial\varepsilon_{\rm f}(=\chi_{\rm v})$ in the vicinity of the QCP in Fig.~\ref{fig:nf_Ef}.

%%%%%
%\textcolor{red}
%{
As discussed in Sect.~2, if we take into account the Al-Au mixed sites, the 4f level at each Yb site has different energy depending on its local environment shown in Fig.~\ref{fig:rate}(a). Furthermore, hybridization paths to the Al-3p states at each Yb site can also be different because of the Al/Au mixed sites as discussed in previous studies~\cite{WM2013,WM2015}. Since the infinite limit of the unit-cell size of the AC corresponds to the quasicrystal, these effects are expected to make many spots of the valence QCP appear in the ground-state phase diagram in Fig.~\ref{fig:PD}, as demonstrated in Ref.~\cite{WM2013}.  Namely, the condensed valence QCPs are considered to be the essential feature specific to the quasicrystal, while the maximum number of valence QCPs are bounded up to 24 at most in the AC as noted in Sect.~2. Hence, by superposing the $\bar{n}_{\rm f}$-$a$ lines in Fig.~\ref{fig:nf_a} and averaging over the configurations of the Al-Au mixed sites and taking the infinite limit of the unit-cell size of the AC, the $\bar{n}_{\rm f}$-$a$ line for the quasicrystal is obtained. Since each configuration is expected to give almost similar $\bar{n}_{\rm f}$-$a$ line shapes to Fig.~\ref{fig:nf_a} (right-decreasing parallel lines in both sides of the QCP and steep decrease at the QCP), the resultant shape is considered to be similar to that shown in Fig.~\ref{fig:nf_a}. 
%}

%%%%%
%\textcolor{red}
%{
Hence, this captures the essential feature of the experimental data, where the quasicrystal Yb$_{15}$Al$_{34}$Au$_{51}$ is located at the position just before the sharp decrease in the Yb valence, as indicated as the QCP in Fig.~\ref{fig:nf_a}~\cite{Nanba2017} and the overall feature is well explained by Fig.~\ref{fig:nf_a}. As noted above, the quasicrystal is considered to be located at the QCP without tuning. Since the $a$ dependence of $\bar{n}_{\rm f}$ shown in Fig.~\ref{fig:nf_a} is expected to be realized for the valence QCP in general, our result calculated in the AC is considered to be relevant for the quasicrystal.  
Thus our calculation supports the experimental finding that the quasicrystal Yb$_{15}$Al$_{34}$Au$_{51}$ is located in the island of the valence QCP. 
%}
%%%%%

%------------------------------------------------------------------------
\section{Summary}

Toward the understanding of the novel quantum critical behavior discovered in the quasicrystal Yb$_{15}$Al$_{34}$Au$_{51}$ and its AC Yb$_{14}$Al$_{35}$Au$_{51}$ under pressure, 
we have discussed the properties of the CEF and the effect of the onsite Coulomb repulsion between the 4f and 5d orbitals at Yb in the AC.  

First, we have analyzed the CEF of the 4f hole at Yb in the AC. 
Because of the low symmetry at the Yb site, the Kramers doublet of the CEF ground state is expressed as the superposition of the $|J_{z}\rangle$ states for $J_{z}=-7/2, -5/2, \cdots, 5/2$, and $7/2$. 
This makes the expectation value of the quadrupole operators non-zero, which provides a possibility that the softening in the elastic constant can appear by the ultrasound measurement.  
The energy of the CEF ground state can be different at each Yb site depending on the configurations of the surrounding atoms due to the Al/Au mixed sites. This reinforces the previous result of the theory of the CVF that the valence QCPs appear as spots in the ground-state phase diagram because of the difference in the effective f-c hybridization at each Yb site due to the Al/Au mixed sites. 

Next, we have constructed the minimal model with the onsite 4f-5d Coulomb interaction $U_{\rm fd}$ on the AC. 
The slave-boson mean-field calculation for the ground state has shown that the valence QCP appears with the intermediate valence of Yb for a reasonable value of $U_{\rm fd}$, 
which is estimated to be a few tenth of eV. 
Furthermore, the 4f-hole number shows the asymmetric lattice-constant dependence around the valence QCP, which captures the essential feature of the recent experiments for the systematic replacement of Al and Au by the other elements such as Ga and Cu respectively in the Yb-Al-Au quasicrystal and AC. 
Our result suggests that the quasicrystal Yb$_{15}$Al$_{34}$Au$_{51}$ is located in the island of the valence QCP.

As the first step of analysis, we have considered the case that the Al/Au mixed sites are occupied by Al in the extended periodic Anderson model. In reality, the 4f level can be different at each Yb site as discussed in Sect.~2 because of the Al/Au mixed sites. 
For more quantitative understanding of the electronic states of the actual quasicrystal and AC, the numerical calculation taking into account the Al/Au mixed sites in the framework based on the model [Eq.~(\ref{eq:EPAM})] is desired in a future study. 
Such a calculation for the 1/1 AC enables us to compare the measured lattice constant dependence of the Yb valence for the AC~\cite{Nanba2017} directly. Furthermore, performing the calculations for the $p/q$ ACs with larger unit cells and analysis of the results from the viewpoint of the size dependence of the unit cell will provide more quantitatively accurate comparison with the data of the quasicrystal by Nanba {\it et al}. 

It is also intriguing to compare our result with the high-pressure measurement~\cite{Watanuki_Yb}. The pressure dependence of the Yb valence in the quasicrystal and AC can be compared with our result in Fig.~\ref{fig:nf_a}. 
The analysis whether the effect of the hydrostatic pressure scales with the effect of chemical substitution to the elements of Al and/or Au by using our model is an interesting subject of  future study.

%-----------------------------------------------------------------------------------------
\section*{Acknowledgment}

The authors thank N. K. Sato for drawing our attention to the recent experimental data and also for allowing us to use the illustration of atoms in Fig.~\ref{fig:Yb_local}. 
One of the authors (S. W.) acknowledges T. Ishimasa for helpful discussion about the symmetry of the quasicrystal and AC. Thanks are also due to M. Matsunami for providing us the photoemission data prior to publication and K. Deguchi, K. Imura, and T. Watanuki for useful discussion about the experimental data. 
This work was supported by JSPS KAKENHI Grant Numbers JP24540378, JP25400369, 
JP15K05177, JP16H01077, and JP17K05555.


\vspace{1cm}

\begin{thebibliography}{99}
\bibitem{Moriya} T. Moriya, {\it Spin Fluctuations in Itinerant Electron Magnetism} (Springer-Verlag, Berlin, 1985); T. Moriya and T. Takimoto, J. Phys. Soc. Jpn. {\bf 64}, 960 (1995).  
\bibitem{Hertz} J. A. Hertz, Phys. Rev. B {\bf 14}, 1165 (1976).
\bibitem{Millis} A. J. Millis, Phys. Rev. B {\bf 48}, 7183 (1993). 
\bibitem{Bauer} E. Bauer, R. Hauser, L. Keller, P. Fischer, O. Trovarelli, J. G. Sereni, J. J. Rieger, and G. R. Stewart, Phys. Rev. B {\bf 56}, 711 (1997). 
\bibitem{Trovarelli} O. Trovarelli, C. Geibel, S. Mederle, C. Langhammer, F. M. Grosche, P. Gegenwart, M. Lang, G. Sparn, and F. Steglich, Phys. Rev. Lett. {\bf 85}, 626 (2000).
\bibitem{Nakatsuji} S. Nakatsuji, K. Kuga, Y. Machida, T. Tayama, T. Sakakibara, Y. Karaki, H. Ishimoto, S. Yonezawa, Y. Maeno, E. Pearson, G. G. Lonzarich, L. Balicas, H. Lee, and Z. Fisk, Nat. Phys. {\bf 4}, 603 (2008).
\bibitem{WM2010} S. Watanabe and K. Miyake, Phys. Rev. Lett. {\bf 105}, 186403 (2010).
\bibitem{Deguchi} K. Deguchi, S. Matsukawa, N. K. Sato, T. Hattori, K. Ishida, H. Takakura, and T. Ishimasa, Nature Mat. {\bf 11}, 1013 (2012). 
\bibitem{Watanuki} T. Watanuki, S. Kashimoto, D. Kawana, T. Yamazaki, A. Machida, 
Y.~Tanaka, and T. J. Sato, Phys. Rev. B {\bf 86}, 094201 (2012). 
\bibitem{Matsukawa2016} S. Matsukawa, K. Deguchi, K. Imura, T. Ishimasa, and N. K. Sato, J. Phys. Soc. Jpn. {\bf 85}, 063706 (2016). 
\bibitem{Deguchi2017} K. Deguchi, invited talk at J-Physics 2017 Sept-25-7.
\bibitem{Matsumoto} Y. Matsumoto, S. Nakatsuji, K. Kuga, Y. Karaki, N. Horie, 
Y. Shimura, T. Sakakibara, A.~H.~Nevidomskyy, and P. Coleman, Science {\bf 331}, 316 (2011).
\bibitem{Ishimasa} T. Ishimasa, Y. Tanaka, and S. Kashimoto, Phil. Mag. {\bf 91}, 4218 (2011).
\bibitem{WM2013} S. Watanabe and K. Miyake, J. Phys. Soc. Jpn. {\bf 82}, 083704 (2013).
\bibitem{WM2015} S. Watanabe and K. Miyake, J. Phys.: Conf. Ser. {\bf 592}, 012087 (2015). 
\bibitem{WM2016} S. Watanabe and K. Miyake, J. Phys. Soc. Jpn. {\bf 85}, 063703 (2016).
\bibitem{WM2014} S. Watanabe and K. Miyake, J. Phys. Soc. Jpn. {\bf 83}, 103708 (2014).
\bibitem{Matsukawa2014} S. Matsukawa,  K. Tanaka,  M. Nakayama,  K.~Deguchi,  K.~Imura,  
H.~Takakura,  S.~Kashimoto,  T.~Ishimasa, and  N.~K.~Sato, J. Phys. Soc. Jpn. {\bf 83}, 034705 (2014). 
\bibitem{Stevens} K. W. H. Stevens, Proc. Phys. Soc. A {\bf 65}, 209 (1952). 
\bibitem{Hutchings} M. T. Hutchings, Solid State Physics {\bf 16}, 227 (1964).
\bibitem{Thalmeier} P. Thalmeirer and B. L\"{u}thi, {\it Handbook on the Physics and Chemistry of Rare Earths} edited by K. A. Gschneider Jr. and L. Erying (Elsevier, Amsterdum, 1991), Vol. 14, p245. 
\bibitem{Kontani} H. Kontani, T. Saito, and S. Onari, Phys. Rev. B {\bf 84}, 024528 (2011).
\bibitem{TKasuya1985} H. Takahashi and T. Kasuya, J. Phys. C: Solid State Phys. {\bf 18}, 2709 (1985).
\bibitem{Andersen1} O. K. Andersen and O. Jepsen, Physica B {\bf 91}, 317 (1977). 
\bibitem{Andersen2} O. K. Andersen, W. Klose, and H. Nohl, Phys. Rev. B {\bf 17}, 1209 (1978). 
\bibitem{Harrison} W. A. Harrison, {\it Electronic Structure and the Properties of Solids} (W. H. Freeman and Co., San Francisco, 1980).  
\bibitem{Read} N. Read and D. M. Newns, J. Phys. C: Solid State Phys. {\bf 16}, 3273 (1983).
\bibitem{OM2000} Y. Onishi and K. Miyake, J. Phys. Soc. Jpn. {\bf 69}, 3955 (2000). 
\bibitem{WM2011} S. Watanabe and K. Miyake, J. Phys.: Condens. Matter {\bf 23}, 094217 (2011).
\bibitem{WMCeRhIn5} S. Watanabe and K. Miyake, J. Phys. Soc. Jpn. {\bf 79}, 033707 (2010).
%\bibitem{Shishido2005} H. Shishido,  R. Settai,  H. Harima, and  Y. $\bar{\rm O}$nuki, J. Phys. Soc. Jpn. {\bf 74}, 1103 (2005). 
\bibitem{Ren2017} Z. Ren, G. W. Scheerer, D. Aoki, K. Miyake, S. Watanabe, and D. Jaccard, Phys. Rev. B {\bf 96}, 184524 (2017). 
%\bibitem{Nanba2017} K. Nanba, K. Deguchi, S. Matsukawa, S. Hirokawa, Y. Yoneyama, K. Imura, and N.~K. Sato, Japan Physical Society autumn meeting (2017) 20pL42-1. 
\bibitem{Nanba2017} K. Imura, H. Yamaoka, S. Yokota, K. Sakamoto, Y, Yamamoto, T. Kawai, K. Namba, S. Hirokawa, K. Deguchi, N. Hiraoka, H. Ishii, J. Mizuki, and  N. K. Sato, in preparation. 
\bibitem{QC_text} P. J. Steinhardt and S. Ostlund, {\it The Physics of Quasicrystals: Lectures and Reprints} (World Scientific, Singapore, 1987).
\bibitem{Watanuki_Yb} T. Watanuki, unpublished.
\bibitem{Jazbec2016} S. Jazbec, S. Kashimoto, P. Ko$\check{\rm z}$elj, S. Vrtnik, M. Jagodi$\check{\rm c}$, Z. Jagli$\check{\rm c}$i$\acute{\rm c}$, and J. Dolin$\check{\rm s}$ek, Phys. Rev. B {\bf 93}, 054208 (2016).  
\bibitem{Das2017} P. Das, P.-F. Lory, R. Flint, T. Kong, T. Hiroto, S. L. Bud'ko, P. C. Canfield, M. de Boissieu, A. Kreyssig, and A. I. Goldman, Phys. Rev. B {\bf 95}, 054408 (2017).
\bibitem{WM2006} S. Watanabe, M. Imada, and K. Miyake, J. Phys. Soc. Jpn. {\bf 75}, 043710 (2006). 
\bibitem{Saiga2008} Y. Saiga, T. Sugibayashi, and D. Hirashima, J. Phys. Soc. Jpn. {\bf 77}, 114710 (2008).
\bibitem{cRPA} F. Aryasetiawan, T. Miyake, and R. Sakuma, {\it The LDA+DMFT approach to strongly correlated materials modeling and simulation} edited by E. Pavarini, E. Koch, D. Vollhardt, and A. Lichtenstein (Forschungszentrum, J\"{u}lich, 2011), Vol.1., Chapt.~7.
\bibitem{Pickett} W. E. Pickett, A. J. Freeman, and D. D. Koelling, Phys. Rev. B {\bf 23}, 1266 (1981). 
\bibitem{Ce_Uff} F. Aryasetiawan, K. Karlsson, O. Jepsen, and U. Sch\"{o}nberger,  Phys. Rev. B {\bf 74}, 125106 (2006).
\end{thebibliography}
\end{document}